\documentclass[
reprint,
superscriptaddress,
showpacs,
nofootinbib,
nobibnotes,
 amsmath,amssymb,
 aps,
prl,
]{revtex4-1}

\usepackage{graphicx}
\usepackage{dcolumn}% Align table columns on decimal point
\usepackage{bm}% bold math
\usepackage{xcolor}
\usepackage{hyperref}

\begin{document}

\title{Electron versus muon neutrino induced cross sections in charged current quasi-elastic processes}

\author{A. Nikolakopoulos}
\email{alexis.nikolakopoulos@ugent.be}
\author{N. Jachowicz}
\email{natalie.jachowicz@ugent.be}
\author{N. Van Dessel}
\affiliation{Department of Physics and Astronomy, Ghent University,\\ Proeftuinstraat 86, B-9000 Gent, Belgium,\\}
\author{K.~Niewczas}
\affiliation{Department of Physics and Astronomy, Ghent University,\\ Proeftuinstraat 86, B-9000 Gent, Belgium,\\}
\affiliation{Institute of Theoretical Physics, University of Wroc{\l}aw \\
Plac Maxa Borna 9, 50-204 Wroc{\l}aw, Poland
}
\author{R.~Gonz\'alez-Jim\'enez}
\author{J.~M.~Ud\'ias}
\affiliation{Grupo de F\'isica Nuclear, Departamento de Estructura de la Materia, F\'isica T\'ermica y Electr\'onica, and IPARCOS,\\
 Universidad Complutense de Madrid, CEI Moncloa, 28040  Madrid, Spain}
\author{V. Pandey}
\affiliation{Center for Neutrino Physics, Virginia Tech, \\
Blacksburg Virginia 24061, USA}

%##########################################################################################################################################

\begin{abstract}
Differences between $\nu_e$ and $\nu_{\mu}$ quasielastic cross sections are essential in neutrino oscillation analyses and CP violation searches for experiments such as DUNE and T2HK. The ratio of these is however poorly known experimentally and for certain kinematic regions theoretical models give contradictory answers.
We use two independent mean-field based models to investigate this ratio using $^{40}$Ar and $^{12}$C targets.
We demonstrate that a proper treatment of the final nucleon's wave function confirms the dominance of $\nu_{\mu}$ over $\nu_e$ induced cross sections at forward lepton scattering. 
\end{abstract}

%##########################################################################################################################################

%\pacs{25.30.Pt, 13.15.+g, 24.10.Jv, 24.10.Cn}

\maketitle

%##########################################################################################################################################
%\section{Introduction}
In recent years, the quest to  elucidate issues concerning neutrino oscillation parameters, the existence of sterile neutrinos, and  CP violation has resulted in a worldwide boom in neutrino experiments and collaborations \cite{NUSTECWP,KatoriMartinireview}.  
Accelerator-based oscillation experiments such as MiniBooNE, T2K, MicroBooNE, and the upcoming DUNE and T2HK facilities \cite{MB:excessPRL, T2K,T2K:CP, DUNE,T2HK, MicroBooNE} rely on neutrino scattering off atomic nuclei in order to detect them in their near and far detectors.
A reliable determination of the neutrino-nucleus cross section is hence pivotal for energy reconstruction and  oscillation analyses.
Theoretical analyses are equally important for dedicated neutrino-nucleus experiments such as e.g. MINER$\nu$A \cite{MINERVA}. 
In view of the determination of oscillation parameters, and in particular the CP-violating phase, an accurate knowledge of  $\nu_e$, $\nu_\mu$ and $\overline{\nu}_e$, $\overline{\nu}_{\mu}$ cross sections over a large kinematic region is indispensable.
The differences between $\nu_e$ and $\nu_{\mu}$ induced  cross sections have been puzzling the community for the last couple of years ~\cite{Ankowski:leptonmass,Martini:Jachowicz,Nieves:2017lij,Day:2012gb}, as it is crucial for the interpretation of the low-energy $\nu_e$ excess~\cite{MB:excessPRL} and for investigations of the neutrino mass hierarchy and the CP-violating phase $\delta_{CP}$~\cite{NUSTECWP,T2K:CP}.

In Ref.~\cite{Martini:Jachowicz} we have shown that the calculated ratio $\sigma_{\nu_e}/\sigma_{\nu_\mu}$  shows important model dependencies. While models agree that electron neutrinos induce larger total cross sections than muon neutrinos \cite{Nieves:2017lij,Ankowski:leptonmass}, the picture can be radically different for specific kinematics. Evaluating cross sections in a mean-field based Hartree-Fock continuum random phase approximation (HF-CRPA) model, we found that for reactions at forward lepton scattering angles, surprisingly charged current muon neutrino-induced interactions show larger cross sections than their electron neutrino counterparts \cite{Martini:Jachowicz}.
In Ref.~\cite{Ankowski:leptonmass} it was argued that a $\nu_{\mu}$ dominance  could be an artifact of an incomplete treatment of the phase space available to the interaction.  This could e.g affect processes studied with a Fermi gas model, as commonly done in experimental analyses.
Meanwhile, the question of which cross section is the larger one at specific kinematics remained unanswered.

In this Letter, we examine the effect of final-state distortion in the modeling of the cross section on this problem
and show that a proper treatment of the distortion of the outgoing nucleon's wave function resolves the issue.  Using two different models and independent codes we  demonstrate that 
describing the reaction with a cross section model that includes nucleon wave functions calculated in a nuclear potential instead of unbound plane waves
reveals that $\nu_\mu$ induced cross sections can indeed be larger than their $\nu_e$  counterparts at forward lepton kinematics. 
Approximating the outgoing nucleon's wave function by a plane wave description,
which introduces inaccuracies related to  
Pauli blocking and orthogonality issues, results in a considerable overestimation of the responses at low energies and forward lepton scattering angles \cite{Megias:2017cuh}.
This kinematic region is especially relevant for the T2K experiment where the oscillated $\nu_\mu$ signal peaks around $300~\mathrm{MeV}$~\cite{T2K:CP} and the low-energy excess of electron-like events found predominantly for forward scattering angles in the MiniBooNE experiment~\cite{MB:excessPRL}.

This Letter is organized as follows: first we show that in  mean-field approaches which include distortion of the final nucleon, the $\nu_\mu$ induced cross section is indeed larger than its $\nu_e$ counterpart in certain kinematic regions.
Then the influence of the kinematics of the process on these cross sections is reviewed, and finally we show that a proper description of Pauli blocking plays an essential role in this effect. 

\begin{figure*}[t]
  \includegraphics[width=0.95\textwidth]{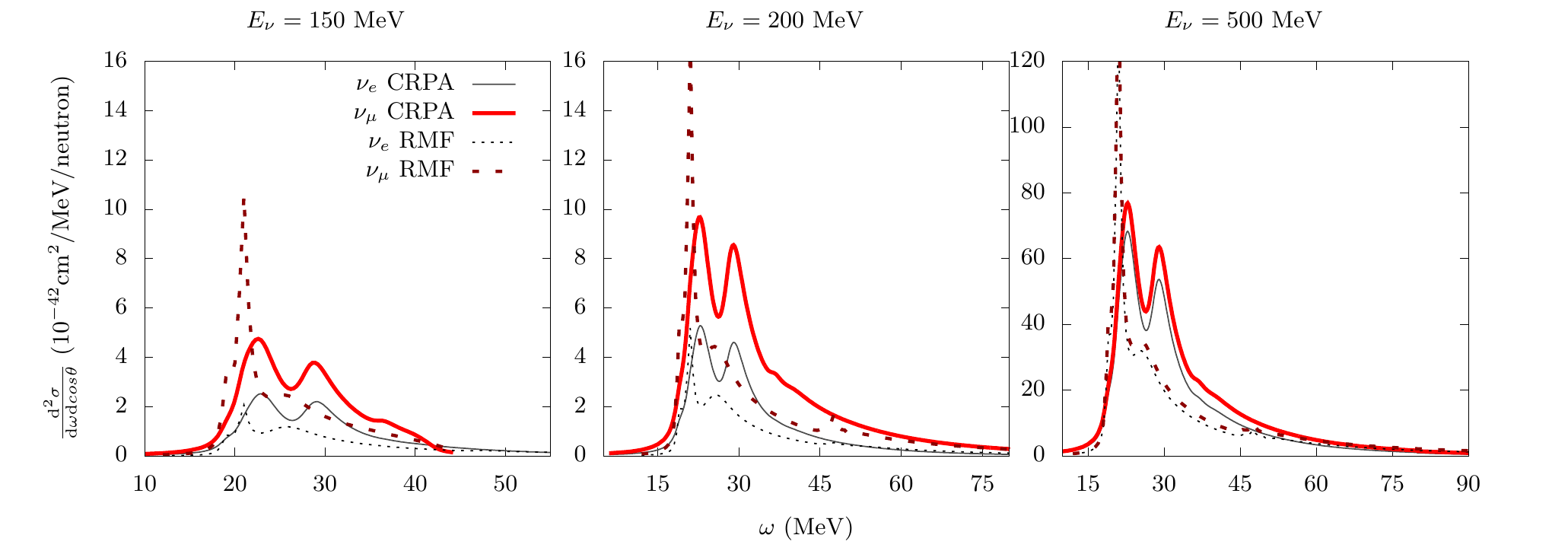}
\caption{CRPA (full lines) and RMF (dashed) cross sections for different incoming neutrino energies for a lepton scattering angle of $\theta_l=5^{\circ}$  for muon (thick lines) and electron neutrino (thin lines) induced interactions with $^{12}$C. }
\label{figRMF}
\end{figure*}

The first approach used in this work, the HF-CRPA model, is based on a mean-field ansatz where the bound-state wave functions are obtained through a Hartree-Fock calculation with a Skyrme interaction \cite{RYCKEBUSCH1988,RYCKEBUSCH1989}.  The final-state nucleon is described with a continuum wave function obtained using the same potential to describe the interaction with the residual system.  Long-range correlations are taken into account through a random-phase approximation approach using the same Skyrme parametrization as residual interaction \cite{CRPAmod, JachoNC,RYCKEBUSCH1989}.  The width of the nucleon states is taken into account in an effective way, folding responses with a Lorentzian  \cite{PRC92}.
In this formalism, the nuclear dynamics are treated in a non-relativistic way, and relativized using the effective procedure proposed in Ref. \cite{Jeschonnek}.  The HF-CRPA approach has been successfully applied to the description of various electroweak scattering processes \cite{RYCKEBUSCH1989, CRPAmod,Pandey:2013cca,PRC92,Pandey:2016,VanDessel:2017ery}.  Short-range correlation effects are very small at the kinematics of interest \cite{Ankowski:leptonmass, VanCuyck:2016}.

The second model is the relativistic mean field (RMF) model for quasielastic one-nucleon knock-out, where the initial and final state nucleon single-particle wave functions are obtained as solutions of the Dirac equation with a mean field potential. The potential is obtained by a self-consistent calculation with a nucleon-nucleon interaction described by a Lagrangian which includes meson fields to parameterize the coupling~\cite{WALECKARMF, RINGRMF}.  It has been shown that the RMF model describes well inclusive electron scattering off nuclei~\cite{PhysRevC.68.048501,PhysRevC.80.024605,PhysRevC.90.035501,Prep}. 

In both approaches, the outgoing wave function is computed in the same nuclear potential as the one used for the bound nucleon states, thereby including the essential feature of orthogonality of  initial and final state.
In this respect, our models contrast with other approaches that ignore secondary interactions of the outgoing nucleon. 

The RMF and HF-CRPA charged current quasi-elastic (CCQE) cross sections for a selected set of  kinematic conditions with small energy and momentum transfers are shown in Fig.~\ref{figRMF}. Although giant resonances cannot be reproduced with the RMF model, its basic features agree well  with  the HF-CRPA results, in particular confirming the $\nu_\mu$/$\nu_e$ ratios found in~\cite{Martini:Jachowicz}, with larger cross sections for the reactions producing the heavier lepton in the final state.  

\begin{figure*}[thb]
\includegraphics[width=0.95\textwidth]{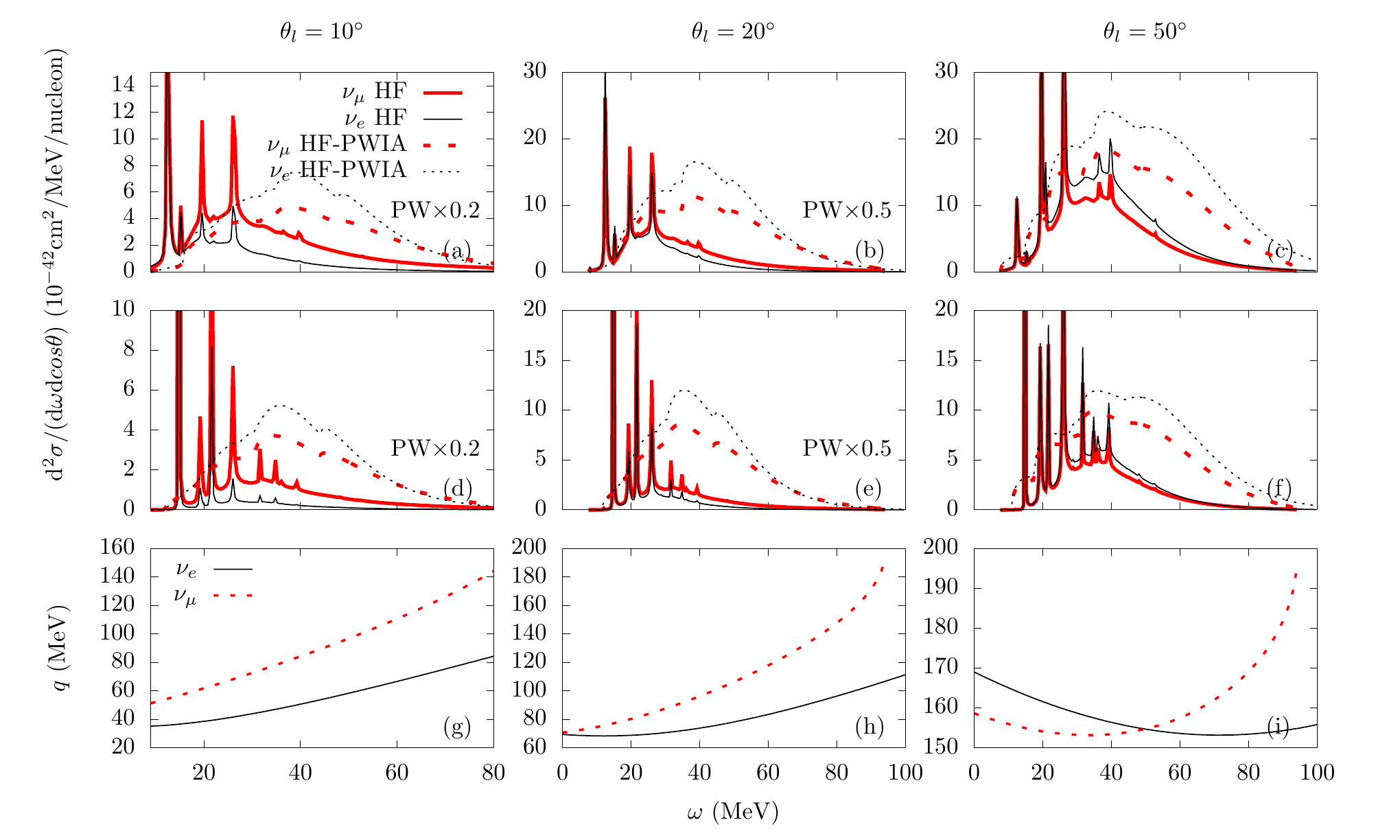}
\caption{HF (full line) and HF-PWIA (dashed) cross section for scattering off $^{40}$Ar for different lepton scattering angles at a fixed neutrino energy of $200~\mathrm{MeV}$, for neutrinos in panels (a)-(c) and antineutrinos in panels (d)-(f).  The momentum transferred to the nucleus as a function of the energy transfer for both muon and electron neutrinos at the same kinematics is shown in panels (g)-(i). To better represent the cross section at the most forward angles the HF-PWIA cross sections have been scaled down in some panels.}
\label{fig:HFPW_argon}
\end{figure*}

The fact that cross sections producing a relatively heavy muon in the final state can be larger than the ones with a light outgoing electron at the same incoming energy, scattering angle and energy transfer,  may seem counter-intuitive, but should in fact not be so surprising.  Looking into the kinematics of the reaction,
 it becomes clear
that for forward lepton scattering angles, at a given energy transfer $\omega$, the corresponding momentum transfer $q$
is larger for muon neutrinos than for electron neutrinos {\em exactly} because the former reaction generates a charged lepton with a
 larger mass.
Indeed, for forward scattering kinematics ($\cos\theta_l \approx 1$), the momentum transferred to the nucleus by a neutrino with incoming energy $E_{\nu}$ that  in the charged-current process transforms into  a lepton with mass $m_l$ and momentum $P_l$, is for fixed  energy transfer $\omega$ given by
\begin{equation}
q =\sqrt{ E_\nu^2 + P_l^2 -2\cos\theta_l E_\nu P_l} \approx E_\nu - \sqrt{\left(E_\nu - \omega \right)^2 - m_l^2 },
\end{equation}
and hence  increases with growing lepton mass.
The struck nucleon receives a smaller momentum $q$ from the electron neutrino than in a muon neutrino-induced interaction.
This brings along larger nuclear responses $R(q,\omega)$ for muon neutrinos.
The lepton kinematic factors that are combined with the responses ~\cite{AmaroResponses, PRC92, Udias:95} to construct the cross section generally tend to favor smaller lepton masses in the forward scattering region, but this effect is not 
large enough to neutralize the dominance of the $\nu_\mu$ responses~\cite{Prep, Jachowicz:JPG2019}.
 This leads to higher cross sections for reactions induced by muon neutrinos, the larger lepton mass in the final state notwithstanding. 
Due to their geometry, near detectors tend to be sensitive mostly to forward lepton scattering events.  The effect observed here is therefore not marginal but could strongly influence total rates
if the angular dependence of the cross section ratio is not fully taken into account.
In this kinematic region the cross section is extremely sensitive to subtle nuclear effects that require careful modeling to be fully understood.
The importance of a meticulous analysis, judging the impact of the different mechanisms at play in the interaction and the nuclear medium, is illustrated in the following paragraphs.

In Fig.~\ref{fig:HFPW_argon}, we demonstrate the link between momentum transfer and cross section, showing the double differential cross section for both electron and muon (anti)neutrino  scattering off argon for a fixed incoming energy, and various scattering angles of the charged lepton together with the momentum transfer in the interaction.
For small momentum transfers the ratio $\sigma_{\nu_\mu}$/$\sigma_{\nu_e}$ can be straightforwardly understood by the difference of momentum transfers depicted in the panels (g) to (i) of the figure.  The cross sections are shown for a Hartree-Fock as well as for a HF plane wave impulse approximation (HF-PWIA) calculation. As described above, in the former one the outgoing nucleon wave is distorted by the presence of the nuclear potential, in the latter one, the wave function is replaced by a plane wave.
Comparing the full HF results to the HF-PWIA ones,  it is obvious that modeling the distortion induced by the nuclear medium is essential.
For small momentum transfers, a plane-wave treatment of the nucleon completely passes by the strong influence of the nuclear potential on the slow final nucleon and yields far too high responses.
\begin{figure*}[t]
\includegraphics[width=0.95\textwidth]{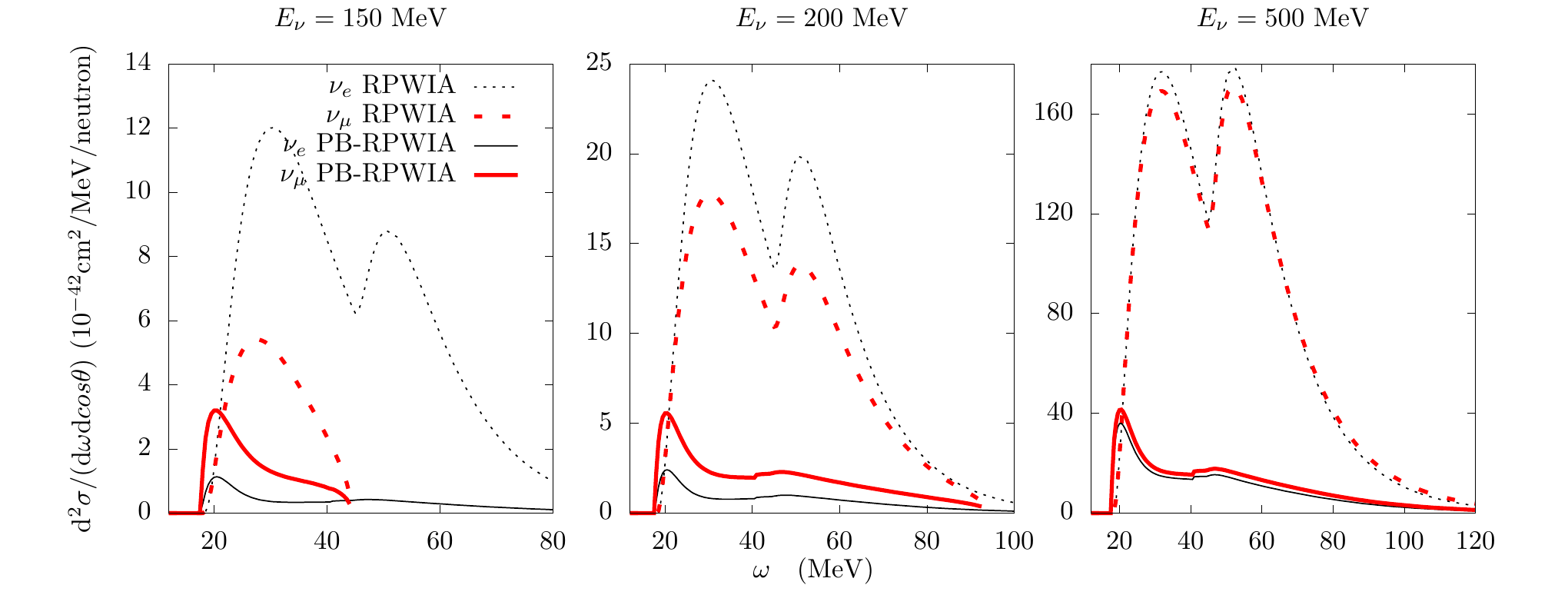}
\caption{Comparison of the $\nu_e$ (thin lines) and $\nu_\mu$ (thick lines) cross sections on carbon for RPWIA (dashed) and PB-RPWIA (solid), in which the non-orthogonal contributions of the plane wave have been projected out. The lepton scattering angle is $\theta_l=5^{\circ}$. }
\label{figPB}
\end{figure*}

In the following, we will compare RMF with relativistic plane wave impulse approximation (RPWIA) results and look into the role of Pauli blocking.
In an RPWIA approach, the outgoing nucleon is modeled by a relativistic plane wave with fixed momentum. On the other hand,
in the RMF approach presented above, the outgoing nucleon has a well-defined energy, but its momentum is only asymptotically defined.
Indeed, only far enough from the nuclear potential, the wave function behaves as an on-shell nucleon with well-defined momentum.
The outgoing wave function in the RPWIA hence has a large component which is non-orthogonal to the bound states of the nucleus.
To illustrate the effect of this on cross sections, we introduce the Pauli-blocked RPWIA (PB-RPWIA).
In this approach the outgoing nucleon wave function is described by a relativistic plane wave which is orthogonalized with respect to the bound states. This is done by projecting out the overlap with the bound states  \cite{Boffi:1982id,CiofidegliAtti:1982lce,Johansson:1999hg,Prep}.

\begin{figure*}
\includegraphics[width=0.95\textwidth]{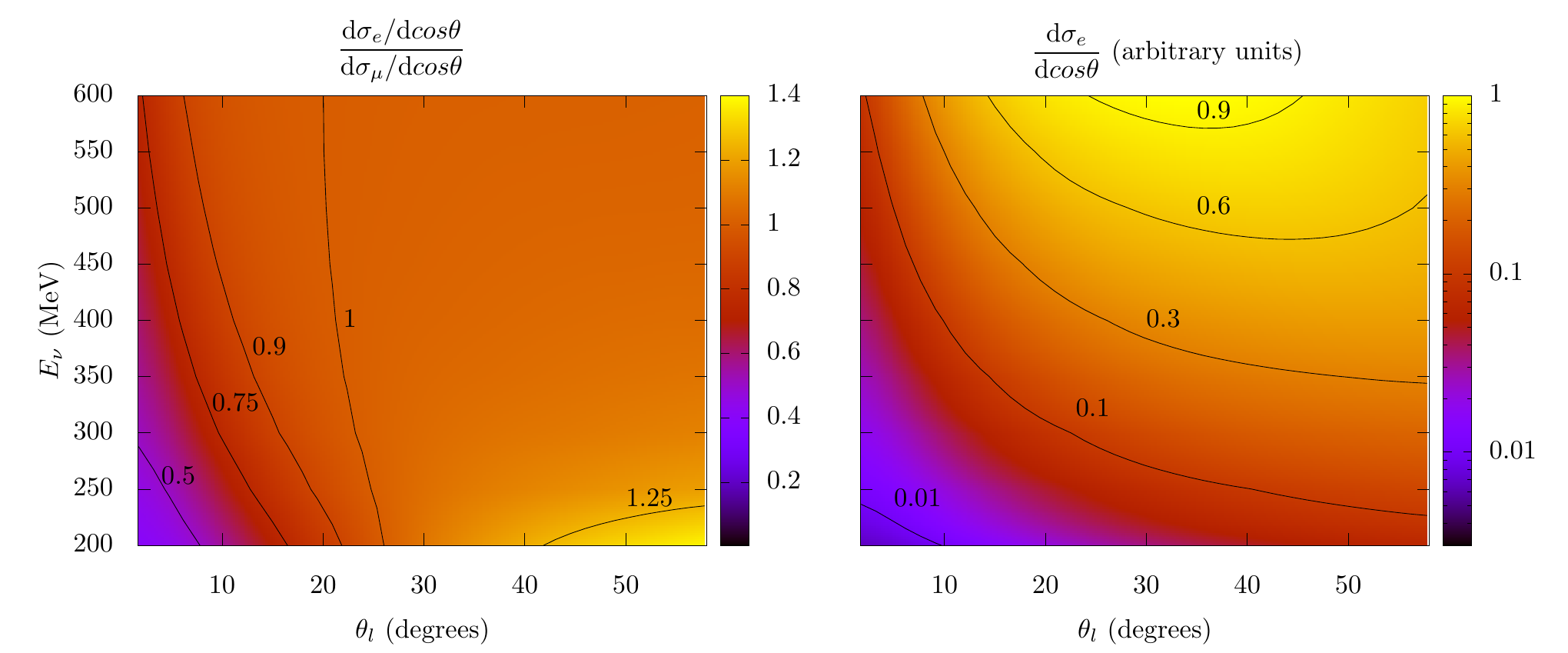}
\caption{Ratio of $^{12}$C cross sections as a function of incoming energy and lepton scattering angle, combined with relative strength of the cross section at the same kinematics (normalized such that the maximum in this kinematic region is 1). Results shown here were obtained within the CRPA approach, RMF ratios are very similar~\cite{Prep}.}
\label{sum}
\end{figure*}

In Fig. \ref{figPB} the RPWIA approach is compared with the PB-RPWIA results for both $\nu_\mu$ and $\nu_e$ induced interactions.
In the RPWIA results the different shells are clearly visible, and the results greatly overestimate the PB-RPWIA cross sections, which reproduce the magnitude of the RMF and CRPA cross sections of Fig~\ref{figRMF}.
It is worth noting that the RPWIA results probe the same kinematic range as the PB-RPWIA, RMF and HF-CRPA approaches. 
In the calculations, no cuts are made in the phase space, the momentum distribution of the initial- and final-state nucleons are completely determined by the wave functions.
As seen in the HF-PWIA results above, in the RPWIA the $\nu_e$ cross sections always have a larger magnitude than the $\nu_\mu$ ones.
However, once the spurious non-orthogonal contributions are removed from the plane wave the situation is reversed for forward kinematics.
In the RMF and HF-CRPA approaches this effect is naturally implemented because the final-state nucleon is constructed from continuum states, with well-defined quantum numbers, in the same potential as the initial state.
In this way Pauli-blocking is implemented in a straightforward quantum mechanical way.
This consistency between initial and final states is also present in the Pauli blocked relativistic Fermi gas model, however the treatment of the nuclear initial and final state as plane waves is unrealistic in this kinematic region.
In the presented approaches nucleons with small momenta can still be emitted from the nucleus, contrary to in a Fermi gas, but the treatment of the final-state wave function naturally leads to a strong reduction of the cross section for slow nucleons.

Figure \ref{sum} summarizes our findings.  The left panel shows the ratio of $\nu_e$ to $\nu_{\mu}$ CRPA cross sections to be smaller than 1 over a large part of the forward scattering region for a broad range of incoming energies.  The right panel testifies that this region represents a considerable part of the scattering strength.

In conclusion, using different models and independently developed codes,
we have shown that taking into account the distortion of the final-state nucleon in the description of charged-current quasi-elastic 
neutrino scattering off atomic nuclei,  muon neutrino-induced CCQE cross sections are larger than the equivalent reaction caused by an electron neutrino for reactions at small energy and momentum transfer.
Indeed, in this kinematic regime the nuclear response is extremely sensitive to  subtle differences in energy and momentum transfer, resulting in sizable differences in cross sections.   
This result sheds light on existing uncertainties in ratios that are essential in the analysis of neutrino oscillation and CP violation searches.\\
As shown, the effect is robust and cannot be seen as an artifact of one model. It is present for neutrinos as well as antineutrinos and manifests itself throughout the nuclear mass table.  It is related to the kinematic peculiarities of the interaction in this regime.  An incomplete treatment of the distortion of the final nucleon's wave function and of Pauli-blocking effects might however obscure the dominance of muon neutrino induced processes over electron neutrino induced ones.
These findings point to the importance of an appropriate description of nuclear effects on neutrino-induced cross sections, especially for forward scattering, and the need for a careful  evaluation of the relevance of various influences of the nuclear medium on the  interaction.

\begin{acknowledgments}
This  work  was  supported  by    the  Research Foundation Flanders (FWO-Flanders), and by the
Special Research Fund,  Ghent University.   The computational resources (Stevin Supercomputer Infrastructure)
and services used in this work were provided by Ghent
University,  the  Hercules  Foundation  and  the  Flemish
Government.   RGJ  was   supported  by  Comunidad de Madrid and UCM under contract No.~2017-
T2/TIC-5252.
KN was partially  supported  by  the  Polish  National
Science  Center  (NCN),  under  Opus  Grant  No.~2016/21/B/ST2/01092, 
as well as by the Polish Ministry of Science and Higher Education, Grant No.~DIR/WK/2017/05.
VP acknowledges the support by the National Science Foundation under grant no.~PHY-1352106.
\end{acknowledgments}

\bibliographystyle{apsrev4-1.bst}
\bibliography{Bibliography-6}% Produces the bibliography via BibTeX.
\end{document}